**Possible origin of the 0.5 plateau in the ballistic conductance of quantum point contacts**

J. Wan, M. Cahay
Department of Electrical and Computer Engineering
University of Cincinnati, Cincinnati OH 45221-0030

P. Debray, R. Newrock
Department of Physics
University of Cincinnati, Cincinnati OH 45221-0030

## Abstract

A non-equilibrium Green function formalism (NEGF) is used to study the conductance of a side-gated quantum point contact (QPC) in the presence of lateral spin-orbit coupling (LSOC). A small difference of bias voltage between the two side gates (SGs) leads to an inversion asymmetry in the LSOC between the opposite edges of the channel. In single electron modeling of transport, this triggers a spontaneous but insignificant spin polarization in the QPC. However, the spin polarization of the QPC is enhanced substantially when the effect of electron-electron interaction is included. The spin polarization is strong enough to result in the occurrence of a conductance plateau at $0.5G_0$ ($G_0 = 2e^2/h$) in the absence of any external magnetic field. In our simulations of a model QPC device, the 0.5 plateau is found to be quite robust and survives up to a temperature of 40K. The spontaneous spin polarization and the resulting magnetization of the QPC can be reversed by flipping the polarity of the source to drain bias or the potential difference between the two SGs. These numerical simulations are in good agreement with recent experimental results for side-gated QPCs made from the low band gap semiconductor InAs.

## I. Introduction

The signature of 1-D ballistic transport is the quantization of its conductance. The conductance exhibits plateaus at integral multiples of $G_0 = 2e^2/h$ [1, 2]. These plateaus are well understood in the framework of the one-electron theory [3]. In 1996, Thomas and co-workers observed an additional plateau at 0.7 $G_0$ in the AlGaAs/GaAs quantum point contact (QPC) in the absence of any magnetic field [4]. Since then, this anomalous plateau, usually referred to as the "0.7 structure", has been observed in both QPCs and relatively long 1D quantum wires [5-9].

To understand its origin, the evolution of the 0.7 structure with temperature, applied magnetic field, and applied voltage has been studied experimentally. Typically, the 0.7 structure exhibits an anomalous temperature dependence showing increase of the conductance to $G_0$ as the temperature is lowered [5]. The zero-bias peak (ZBA) has been observed in the non-linear differential conductance and the appearance of the ZBA peak is related to the disappearance of the 0.7 structure [10]. In a parallel magnetic field, the 0.7 structure evolves smoothly into the Zeeman spin-split $0.5G_0$ plateau [4]. At the same time, the actual position of the 0.7 plateau may vary in the range [0.5 to 0.7] $G_0$ depending on the electron density, the channel length and the lateral confining potential [11, 12].

The 0.7 structure has also generated intense theoretical efforts and several models have been proposed, but at present there seems to be no consensus on its origin. The Kondo model successfully predicts the temperature dependence and the ZBA for the 0.7 structure [13, 14], but it contradicts with the recent observation of the static spin polarization in a hole QPC [10]. Based on strictly one-dimensional Luttinger liquid state, a Wigner-cystal model has been used to explain the 0.7 structure and its temperature dependence [15]. However, a QPC is not necessarily a strictly 1D system, so the Luttinger liquid theory may not be applicable here. Another model links the 0.7 structure to the spontaneous spin polarization in the QPC [16-19]. Since the static spin polarization has been experimentally found to be associated with the 0.7 structure [10], it deserves our attention. The local exchange energy can introduce a spin-split energy gap. When Fermi energy lies in the spin-split gap, a static spin polarization gives a plateau in the range of [0.5 to 0.7] $G_0$. However, all these spin polarization models need to implant magnetic impurities or apply an initial small magnetic field [16-19] to trigger initial spin imbalance in the simulations.

In a recent report [20], we presented experimental evidence of the 0.5 plateau in side-gated quantum QPCs, when the confining potential of the QPC is made sufficiently asymmetric. In these structures, the asymmetrical lateral spin orbit coupling (LSOC) is used to introduce the initial spin unbalance [20]. The QPCs were made from InAs (with a large intrinsic SOC, more than one order of magnitude larger than GaAs [21]) quantum-well (QW) structures with a two-dimensional electron gas (2DEG) in the well. Figure 1(a) is the SEM of a QPC device made on InAs QW structure. The QPC channel is created by negatively biasing the two side gates (SG) UG and LG. The confining potential of the QPC is controlled by the bias voltages of the side gates and can be made asymmetric by applying unequal voltages to these gates. Figure 1(b) shows a representative plot of measured conductance at 4.2K of the InAs QPCs as a function of common-mode side gate voltage in the absence of any applied magnetic filed. An additional short plateau at conductance $G \sim 0.5$ $G_0$ was observed when the confining potential was made highly aymmetic (AS). It was absent in the symmetric (S) case. Such a plateau is known to result due to Zeeman spin splitting of the lowest 1D subband in an applied magnetic field. This happens when the Fermi level lies in the spin-split energy gap. The occurrence of a 0.5 plateau in the ballistic conductance is a signature of complete spin polarization. The observed 0.5 structure of Fig. 1(b) indicates spontaneous spin polarization of the conduction electrons. It was observed only when the transverse confining potential of the QPC was made asymmetric by appropriately adjusting the side-gate voltages. The 0.5 plateau was also observed when the confinement asymmetry was reversed by flipping the asymmetry of the gate voltages.

In ref.[20], a simple analytical model of the potential energy profile felt by the electrons in the central portion of the QPC was used in conjunction with a non-equilibrium Green function formalism (NEGF) to provide a theoretical explanation of the experimentally observed 0.5 plateau of Fig. 1(b). We showed that the 0.5 plateau is related to a spontaneous spin polarization induced by LSOC in a side-gated QPC. More precisely, the 0.5 $G_0$ conductance plateau appears in QPCs in the absence of any external magnetic field as a result of three ingredients: an asymmetric lateral confinement, a LSOC, and a strong electron-electron interaction. The asymmetry in the LSOC is required to trigger a small initial spin imbalance. This is true even in a single-electron description of carrier transport through the QPC [22]. However, the small spin imbalance is strongly enhanced when the effect of electron-electron interaction is taken into account to a point that near-perfect spin polarization in the QPC is reached resulting in the 0.5 plateau in the conductance. In this paper, we investigate in more detail the nature of the 0.5 plateau by considering more realistic potential energy profiles in side-gated QPCs using a self-consistent NEGF that includes the effect of space-charge in the structures. We study the effects of temperature and strength of the electron-electron interaction on the shape of the 0.5 conductance plateau. This paper is organized as follows. In the next section, we discuss in detail the origin of the 0.5 plateau. In section III, we describe the NEGF approach used to calculate the charge densities of spin-up and spin-down electrons throughout the QPC. Section IV gives the results of our numerical simulations. Finally, Section V contains our conclusions.

## II. The Origin of the 0.5 Plateau

We model the QPC using the configuration shown in Fig. 2 where the white region represents the mesa etched quantum wire with openings on either side. The gray areas represent the etched isolation trenches that define the dimensions of the QPC. There are four contacts connected to the QPC device, source, drain and two SGs. Symmetric and asymmetric SG voltages can be applied. Since the QPC of Fig. 1(a) is made from a nominally symmetric InAs QW, spatial inversion asymmetry was assumed to be negligible along the growth axis ($z$ axis) of the QW and the corresponding Rashba spin-orbit interaction was neglected. The Dresselhaus spin-orbit interaction due to the bulk inversion asymmetry in the direction of current flow is also neglected. The only spin-orbit interaction considered is the LSOC due to the lateral confinement of the QPC channel, provided by the isolation trenches and the bias voltages of the side gates. To understand the origin of the 0.5 plateau, we begin with the single-particle Hamiltonian, which is given by

$$\begin{aligned} H &= H_0 + H_{SO} \\ H_{SO} &= \beta\vec{\sigma}\cdot(\vec{k}\times\vec{\nabla}U) = \vec{\sigma}\cdot\vec{B}_{SO} \end{aligned} \quad (1)$$

where $H_0 = \frac{1}{2m^*}\left(p_x^2 + p_y^2\right) + U(x,y)$, $\beta$ is the intrinsic SOC parameter, $\vec{\sigma}$ is the vector of Pauli spin matrices, and $\vec{B}_{SO}$ is the effective magnetic field, which is induced by the LSOC. The 2DEG is assumed to be located in the (*x*, *y*) plane, *x* being the direction of current flow from source to drain and *y* the direction of confinement of the channel. *U*(x, y) is the confinement potential, which includes the potential introduced by contact gates and conduction band discontinuity $\Delta E_c(y)$ at the InAs/air interface.

Figure 3(a) gives a schematic representation of the confining potential along y direction when a symmetric side-gate voltage is applied. The effective magnetic field $\vec{B}_{SO}$ has exactly the same magnitude but opposite directions at the opposite transverse edges of the QPC. Moving electrons with opposite spins experience opposite SOC forces that leads to an accumulation of opposite spins at the opposite transverse edges. The spin-up is the majority spin species on edge I of the QPC and the minority species on edge II. The difference of spin density is anti-symmetric about $y = w_1/2$ giving zero net polarization.

When an asymmetric SG voltage is applied on the QPC, the potential profile changes from the symmetric dashed line to the asymmetric full line as shown in Fig. 3(b). The spin-up population on the left edge I exceeds the spin-down one on the right edge II. This results in a net spin-up polarization, which is the initial imbalance between spin-up and spin-down electrons induced by the asymmetric LSOC. It can be shown that when the asymmetry between $V_{sg1}$ and $V_{sg2}$ is reversed, the direction of spin polarization is reversed. The strong repulsive Coulomb e-e interaction enhances this imbalance. As a result, the spontaneous spin polarization can reach nearly 100% in the regime of single-mode transport and the 0.5 conductance plateau can result.

## III. The NEGF Model for Side-Gated QPC

The conductance through the QPC was calculated using a NEGF method under the assumption of ballistic transport [23]. Within the NEGF, the Green function associated to the QPC is then a ($2N_xN_y \times 2N_xN_y$) matrix (where $N_x$ and $N_y$ are the number of grid points along and perpendicular to the direction of current flow and the factor two is needed to distinguish the degree of spin polarization) and is given by

$$G(E) = (EI - H - \Sigma_S - \Sigma_D - \Sigma_{\text{int}})^{-1}, \quad (2)$$

where $\Sigma_S$ and $\Sigma_D$ are the self-energy terms representing the coupling of the source and drain contacts [23] and $\Sigma_{\text{int}}$ is the electron-electron interaction self-energy. We used a Hartree-Fock approximation following Lassl et al. [19] to include the effects of Coulomb e-e interaction in the QPC. In this approach, the interaction between two electrons located at $(x, y)$ and $(x', y')$ is modeled using the contact potential

$$V_{\text{int}}(x, y; x', y') = \gamma\delta(x - x', y - y'), \tag{3}$$

where $\gamma$ is the interaction strength equal to a few times of $\hbar^2 / 2m^*$.

The interaction self-energy with spin σ is then given by

$$\Sigma_{\text{int}}^{\sigma}(x, y) = \gamma n_{-\sigma}(x, y), \tag{4}$$

where $n_{-\sigma}(x, y)$ is the density of electrons with spin -*σ*.

The interaction self energy $\Sigma_{\text{int}}^{\sigma}(x, y)$ is different for the two spin populations injected from the contacts. A spin-up electron encounters a potential which is proportional to the density of spin-down electrons, and vice versa. This leads to a repulsive interaction between electrons with opposite spin directions. Any external source leading to an imbalance between the density of spin-up and spin-down electrons is increased by the addition of the self-energy term $\Sigma_{\text{int}}^{\sigma}(x, y)$. In our case, it is the asymmetric LSOC, which leads to the initial imbalance.

Once *H*, $\Sigma_S$, $\Sigma_D$ and $\Sigma_{\text{int}}$ are known, the Green function (*G*) can be calculated from Eq. (2) and all the other quantities of interest can be found out from the following set of equations.

1) The broadening matrices for the source and drain contacts

$$\Gamma_S(E) = i(\Sigma_S - \Sigma_S{}^{\dagger}), \;\; \Gamma_D(E) = i(\Sigma_D - \Sigma_D{}^{\dagger}). \tag{5}$$

2) The spectral functions of the source and drain contacts

$$A_S(E) = G\Gamma_S(E)G^{\dagger}, \;\; A_D(E) = G\Gamma_D(E)G^{\dagger}. \tag{6}$$

3) The density matrix

$$\rho = \int_{-\infty}^{\infty} \frac{A_S(E)f_S(E) + A_D(E)f_D(E)}{2\pi} dE, \tag{7}$$

where $f_S$ and $f_D$ are Fermi-Dirac distributions in the source and drain contacts, respectively.

5) The transmission coefficient

$$T(E) = Trace[\Gamma_S G \Gamma_D G^{\dagger}]. \tag{8}$$

6) The current through the QPC under the assumption of ballistic transport

$$I = \frac{q}{h}\int_{-\infty}^{\infty} T(E)\left[f_S(E) - f_D(E)\right] dE \cdot \qquad (9)$$

Based on Eq.(5), Eq.(6) and Eq.(7), we can get the density matrix $\rho$ from the Green function matrix $G$. The diagonal elements of the density matrix determine the spin-up density $n_\uparrow$ and spin-down density $n_\downarrow$ and hence the interaction self-energy $\sum_{\text{int}}$ from Eq.(4). Since the interaction self-energy depends on the Green function matrix and vice-versa, an iterative procedure is required to obtain the final results.

## IV. Numerical Results

We consider a side-gated QPC made from InAs QW structure with a 2DEG in the well. The low band gap semiconductor InAs has a large intrinsic SOC. The effective mass in the InAs channel was set equal to $m^*= 0.023m_0$, where $m_0$ is the free electron mass. Unless otherwise stated, all calculations were performed at a temperature $T = 4.2\ K$. Following Lassl et al. [19], the strength of the parameter $\gamma$ in the interaction self energy was set equal to $3.7\hbar^2/2m^*$. The strength of the parameter $\beta$ in the LSOC was set equal 200 $\text{Å}^2$ [20]. Unless otherwise stated, the geometrical parameters $l_1$, $l_2$, $w_1$ and $w_2$ were selected to be 68, 36, 48 and 16 nm, respectively. These parameters are smaller than the experimental values of the QPC shown in Fig.1 (a) and were chosen to reduce computational time. The potential at the source was set equal to 0V and the one at the drain $V_d$ to 0.3 mV in all simulations. An asymmetry in the potential of the SGs was introduced by taking $V_{sg1}$ = 0.2 V + $V_{sweep}$ and $V_{sg2}$ = -0.2 V + $V_{sweep}$ and the conductance of the constriction was studied as a function of the sweeping (or common mode) potential, $V_{sweep}$. The Fermi energy was equal to 106.3 meV in the source contact and 106meV in the drain contact, ensuring single-mode transport through the QPC.

At the interface between the rectangular region of size $w_2 \times l_2$ and vacuum, the conduction band discontinuity was modeled as

$$\Delta E_c(y) = \frac{\Delta E_c}{2}\left[1 + \cos\frac{\pi}{d}\left(y - \frac{w_1 - w_2}{2}\right)\right] \qquad (10)$$

at the bottom interface, and

$$\Delta E_c(y) = \frac{\Delta E_c}{2}\left[1 + \cos\frac{\pi}{d}\left(\frac{w_1 + w_2}{2} - y\right)\right], \qquad (11)$$

at the top interface to achieve a smooth conductance band change, where $d$ was selected to be in the nm range to represent a gradual variation of the conduction band profile from the inside of the quantum wire to the vacuum region. A similar grading was also used along the walls going from the wider portion of the channel to the central portion of the QPC, as shown in Fig.2. The gradual change in $\Delta E_c(y)$ is responsible for the LSOC triggering the spin polarization of the QPC in the presence of an asymmetry between $V_{sg1}$ and $V_{sg2}$, as discussed below. The parameter $d$ appearing in Eqns. (10) and (11) was set equal to 1.6 nm. Similar results were obtained when $\Delta E_c(y)$ was linearly changed at the interface and $d$ was set equal to 0.8 nm and 1.2 nm. The conductance of the QPC was then calculated using the NEGF outlined in section II using a non-uniform grid configuration containing more grid points at the interface of the QPC with vacuum [24].

Figure 4 is our main result which shows plots of the conductance of the QPC as a function of $V_{sweep}$ for symmetric ($\Delta V_{SG} = V_{sg1} - V_{sg2} = 0$) and asymmetric ($\Delta V_{SG} = V_{sg1} - V_{sg2} = 0.4$ V) confinements. The dashed curve is the conductance calculated with the symmetric confinement and only one plateau at $2e^2/h$ is observed. The oscillation in the conductance for $V_{sweep} > 0$ is a result of multiple reflections between the ends of the central rectangular portion of the QPC. The full curve labeled "$G_\uparrow + G_\downarrow$" is the conductance with asymmetric confinement and clearly indicates the presence of a plateau in conductance around $e^2/h$ besides the normal $2e^2/h$ plateau. Not shown here, the contributions of $G_\downarrow$ and $G_\uparrow$ to the conductance was found to be switched when the polarity of $\Delta V_{sg}$ was flipped. Figure 4 also shows the contribution of the majority and minority spin bands as a function of $V_{sweep}$. $G_\uparrow$ and $G_\downarrow$ are found to be decreasing and increasing, respectively, near $V_{sweep} = 0$ V leading to a negative differential region in $G_\uparrow + G_\downarrow$. This feature is quite common in the numerical simulations of QPCs [16,17,19,25]. It is partly due to the effect of multiple reflections at the edges of the narrow portion of the QPC but also depends very strongly on the exact shape of the potential energy landscape in the QPC [16]. As shown in Fig.5, the size of the conductance modulation Δ on the plateau gets smaller when the aspect ratio $w_2/l_2$ is closer to unity, in agreement with the experimental results shown in Fig.1 for which $w_2/l_2 \sim 1$.

Figure 6 shows that the 0.5 plateau is rather sensitive to the choice of the parameter γ, but is otherwise robust. If the constriction of the QPC is narrow enough for transport to be single-mode, the 0.5 plateau is fairly robust as a function of temperature, as shown in Fig.7. Even at a temperature of 40 K, a remnant of the 0.5 plateau can still be seen. The difference in conductance $G_\uparrow$ and $G_\downarrow$ in the transition region from the cut off region and the first quantized plateau $2e^2/h$ can be understood by plotting the potential energy profile for the majority and minority spin bands. Figure 8 shows a plot of potential energy profile $U_\uparrow$ and $U_\downarrow$ in the central portion of the QPC for $V_{sweep} = 0$ V, where $U_\uparrow(x, y) = U(\mathrm{x, y}) + \sum_{int}^{\uparrow}$ and $U_\downarrow(x, y) = U(\mathrm{x, y}) + \sum_{int}^{\downarrow}$. The potential energy profile $U_\downarrow$ has a camelback shape which prevents the flow of minority spins through the QPC, whereas $U_\uparrow$ has a saddle shape. The Fermi level is then in between the local maximum (point A in Fig.8(a)) and local minimum (point B in Fig.8(b)) felt by the minority and majority band, respectively. The majority spin band is propagating through the channel but the minority spin band is evanescent. The difference in potential energy between points A and B located at the center of the rectangular portion of the QPC is equal to 0.033eV which is about 100 times $k_BT = 0.36$ meV for $T = 4.2$ K. This explains why the 0.5 plateau can still be observed at temperature as high as 40K in Fig.7. With further increase in $V_{sweep}$, the Fermi level is now above both maxima of the potential energy profile $U_\uparrow$ and $U_\downarrow$ in the central portion of the QPC, both channels are propagating and the difference between $U_\uparrow$ and $U_\downarrow$ disappears gradually. The conductance then reaches a maximum of $2e^2/h$.

A plot of $n_\uparrow$(x,y) - $n_\downarrow$(x,y) for $V_{sweep} = 0$ is shown in Fig. 9 to illustrate that the spin imbalance is the strongest in the central (rectangular) portion of the QPC. To better illustrate the efficiency of spin injection through the QPC, we have plotted in Fig.10 the spin polarization $\alpha(x,y) = (n_\uparrow(x,y) - n_\downarrow(x,y))/(n_\uparrow(x,y) + n_\downarrow(x,y))$. Its maximum value was found to be 81% but it can be tuned closer to 100 % by changing the QPC structural parameters. For comparison, we plot in Figures 11(a) and 11(b) the quantities $n_\uparrow$(x,y) - $n_\downarrow$(x,y) and α(x,y) when the parameter γ is set equal to zero, i.e., the effects of electron-electron interaction are neglected. In this single electron picture of carrier transport, the imbalance induced by the asymmetry in LSOC triggers an accumulation of opposite species on the opposite edges of the central portion of the QPC [22]. This spin imbalance leads to an asymmetry in α(x,y) along the width of the channel, as shown in Fig.11(b). In this figure, the maximum of α(x,y) is about a factor 400 below the maximum observed in Fig.10. This stresses the importance of electron-electron interaction in the channel due to the substantial difference between the potential energy profiles $U_\uparrow$ and $U_\downarrow$ acting on the two different spin species in the central portion of the QPC.

In summary, NEGF simulations of spontaneous spin polarization in side-gated QPCs present the following picture. Below threshold, the Fermi level in the source is below the minimum of the potential energy profile for either spin band and the conductance is close to zero. When the confining potential of the QPC is symmetric, opposite spin accumulations induced by LSOC at the transverse edges cancel each other and there is no net spin polarization. When the confinement potential is made asymmetric by applying a small potential difference between the two side gates, the asymmetric LSOC triggers a small imbalance between the majority and minority spin bands. When this imbalance is fed back into the local e-e interaction self-energy term, the difference in the potential energy profiles felt by the two different spin channels becomes quite drastic [26]. In addition, we have shown that the shape of the anomalous plateau is quite robust but otherwise sensitive to the value of the parameter γ in Eq.(4), i.e., the exact form used to model the effect of exchange-correction potential energy. Extensive conductance measurements around the 0.5 plateau of QPC could therefore be used to refine the details of the theoretical treatment and understand better the importance of e-e interaction.

## V. Conclusion

The use of asymmetrically biased lateral QPCs offers an all-electrical way to generate highly spin-polarized current avoiding the need for ferromagnetic contacts or external magnetic field. The use of two QPCs in series with a submicron long channel in between whose width could be tuned by two additional side gates could therefore pave the way for the first demonstration of an all-electrical realization of the Datta-Das SpinFET. The same approach could provide an all electrical means to realize semiconducting quantum computing gates.

## Acknowledgment

This work was supported by NSF award ECCS 0725404.

## Figure Captions

Figure 1(a) Scanning electron micrograph (SEM) of a side-gated InAs QPC. SG1 and SG2 are side gates. Dark areas are trenches cut by wet etching. Light areas are the wafer surface with the 2DEG underneath. (b) Conductance of InAs QPC at 4.2K in the absence of applied magnetic filed. The 0.5 conductance plateau is observed when the confining potential is asymmetric (AS). It is absent when the confining potential is symmetric (S) [20].

Figure 2: Schematic of the QPC configuration used in the numerical simulations. The geometrical dimensions $l_1$, $l_2$, $w_1$ and $w_2$ are selected to be 68, 36, 48 and 16nm, respectively. In all simulations, $V_s$ = 0V, $V_d$ = 0.3mV.

Figure 3: (a) Schematic representation of the confining potential energy along y direction at x = $l_1/2$ when symmetric SG voltages are applied; (b) Same for asymmetric SG voltages.

Figure 4: Conductance of a QPC as a function of $V_{sweep}$ with the geometrical dimensions shown in Fig.2. The SG biasing parameters for the symmetric confinement case are $V_{sg1}$ = $V_{sg2}$ = $V_{sweep}$; for the asymmetric case, $V_{sg1}$ = 0.2V + $V_{sweep}$ and $V_{sg2}$ = -0.2V + $V_{sweep}$. The temperature is set equal to 4.2K. The top dashed curve corresponds to the symmetric confinement. The individual contributions of the two spin bands to the full curve "$G_\uparrow + G_\downarrow$" are is shown when asymmetric confinement is applied in the simulations. The 0.5 plateau is clearly visible then.

Figure 5: Conductance of a QPC as a function of $V_{sweep}$ close to the 0.5 plateau for different choices of the aspect ratio $w_2/l_2$. The parameter γ is set equal to 3.7 in units of $\hbar^2/2m^*$. $V_s$ = 0V, $V_d$ = 0.3mV, $V_{sg1}$ = 0.2V + $V_{sweep}$, $V_{sg2}$ = -0.2V + $V_{sweep}$. The temperature is 4.2K. The geometrical dimensions $l_1$, $w_1$, and $w_2$ are selected to be 68, 48 and 16nm, respectively. From left to right, the curves correspond to $l_2$ equal to 36, 32, 26, and 24 nm, respectively. The corresponding value of the conductance modulation Δ on the plateau is equal to 0.29, 0.24, 0.17, and 0.12 ($2e^2/h$), respectively.

Figure 6: Dependence of the 0.5 conductance plateau on parameter γ (in units of $\hbar^2/2m^*$). The biasing parameters of the QPC are $V_s$ = 0V, $V_d$ = 0.3mV, $V_{sg1}$ = 0.2V + $V_{sweep}$, $V_{sg2}$ = -0.2V + $V_{sweep}$. The temperature is 4.2K. The geometrical dimensions $l_1$, $l_2$, $w_1$ and $w_2$ are selected to be 68, 36, 48 and 16nm, respectively.

Figure 7: Temperature variation of the 0.5 conductance plateau for a QPC with the geometrical configuration shown in Fig.2. The biasing parameter of the QPC are $V_s$ = 0V, $V_d$ = 0.3mV, $V_{sg1}$ = 0.2V + $V_{sweep}$, $V_{sg2}$ = -0.2V + $V_{sweep}$. The parameter γ is set equal to 3.7 in units of $\hbar^2/2m^*$. The geometrical dimensions $l_1$, $l_2$, $w_1$ and $w_2$ are selected to be 68, 36, 48 and 16nm, respectively.

Figure 8: (a) Potential energy profile for minority spin band in the central portion of QPC. (b) Same for majority spin band. The parameter γ is set equal to 3.7 in units of $\hbar^2/2m^*$, $V_{sweep}$ = 0V, and the temperature $T$ = 4.2K. The geometrical dimensions $l_1$, $l_2$, $w_1$ and $w_2$ are selected to be 68, 36, 48 and 16nm, respectively. There is a difference of 33.6meV between the points labeled A and B.

Figure 9: Spatial variation of the difference of the density for majority and minority spins across the QPC for $V_s$ = 0V, $V_d$ = 0.3 mV, $V_{sg1}$ = 0.2V, $V_{sg2}$ = -0.2V (i.e., $V_{sweep}$ = 0V). The temperature $T$ = 4.2K and γ = 3.7 in units of $\hbar^2/2m^*$. The geometrical dimensions $l_1$, $l_2$, $w_1$ and $w_2$ are selected to be 68, 36, 48 and 16nm, respectively.

Figure 10: Spatial dependence of the spin polarization $\alpha(x,y) = \left(n_\uparrow(x,y) - n_\downarrow(x,y)\right)/\left(n_\uparrow(x,y) + n_\downarrow(x,y)\right)$ in the central portion of the QPC for $V_s$ = 0V, $V_d$ = 0.3mV, $V_{sg1}$ = 0.2V, $V_{sg2}$ = -0.2V (i.e., $V_{sweep}$ = 0V). The temperature $T$ = 4.2K and γ = 3.7 in units of $\hbar^2/2m^*$. The geometrical dimensions $l_1$, $l_2$, $w_1$ and $w_2$ are selected to be 68, 36, 48 and 16nm, respectively.

Figure 11: Plot of (a) the spatial variation of the difference of the density for majority and minority spins across the QPC and (b) the spatial dependence of the spin polarization $\alpha(x,y) = \left(n_\uparrow(x,y) - n_\downarrow(x,y)\right)/\left(n_\uparrow(x,y) + n_\downarrow(x,y)\right)$ in the central portion of the QPC when the effects of the electron-electron interaction in the QPC are neglected (γ = 0). The biasing conditions, physical and device parameters are the same as in Figures 9 and 10.

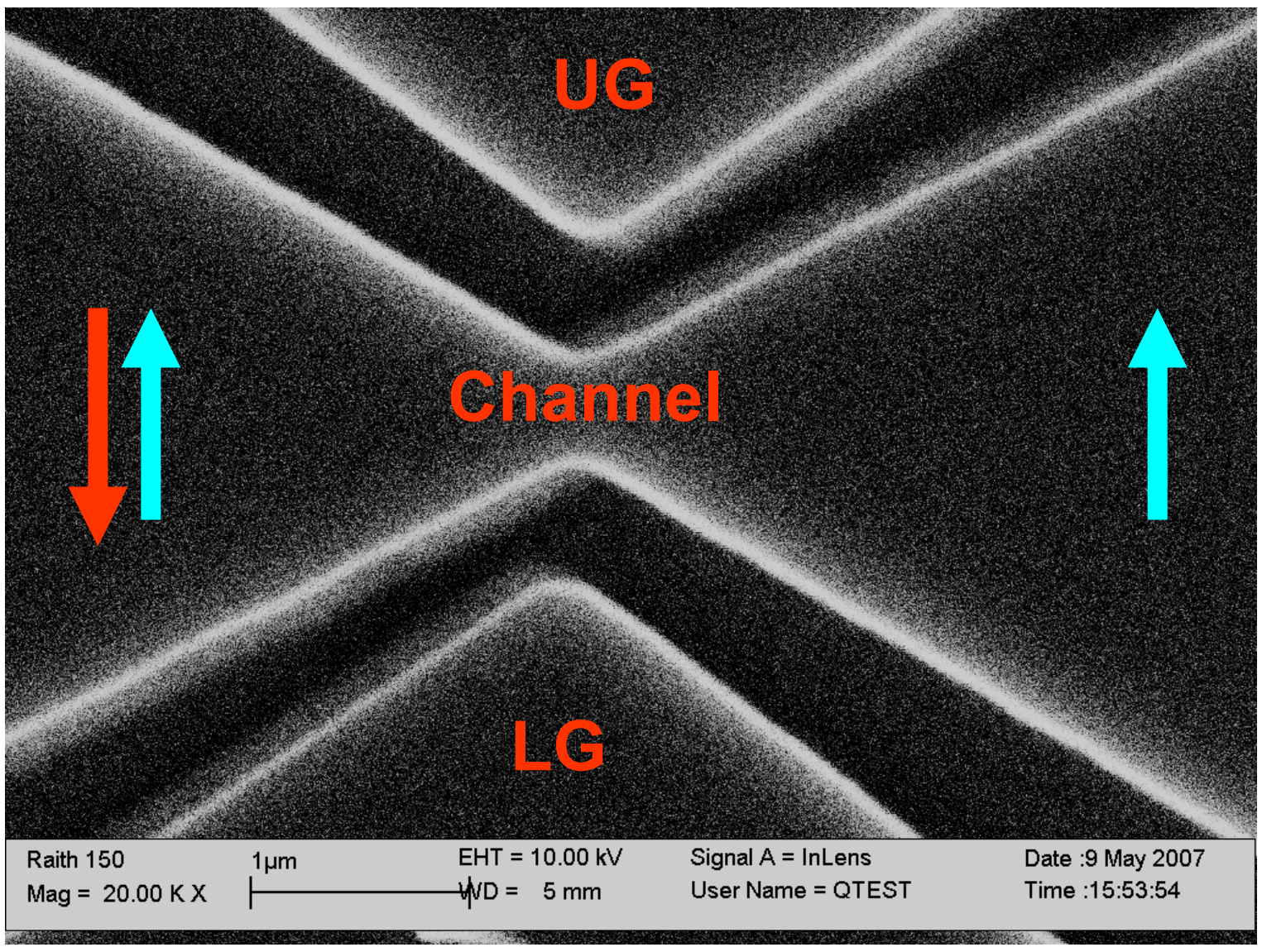

Figure 1(a)

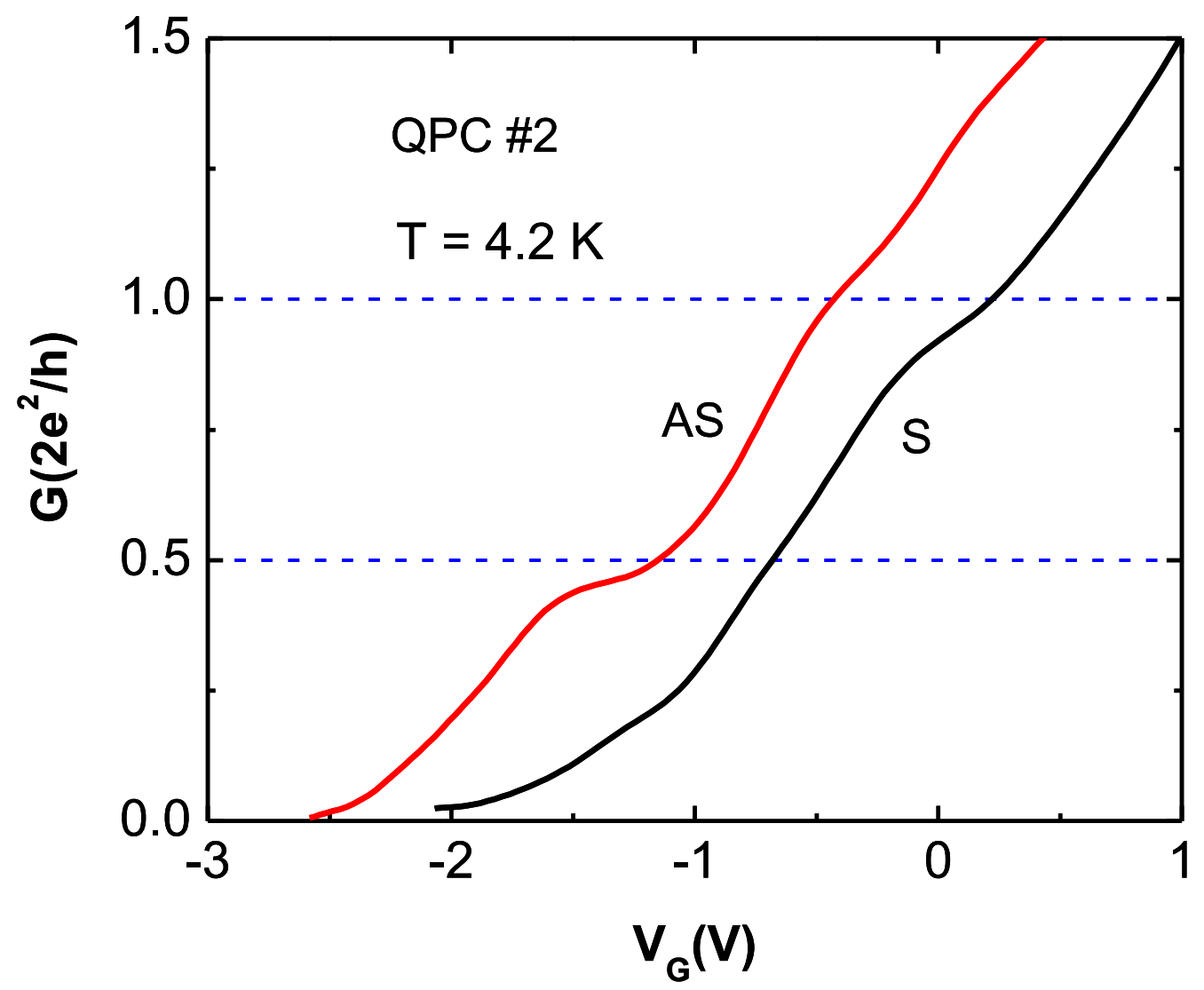

Figure 1(b)

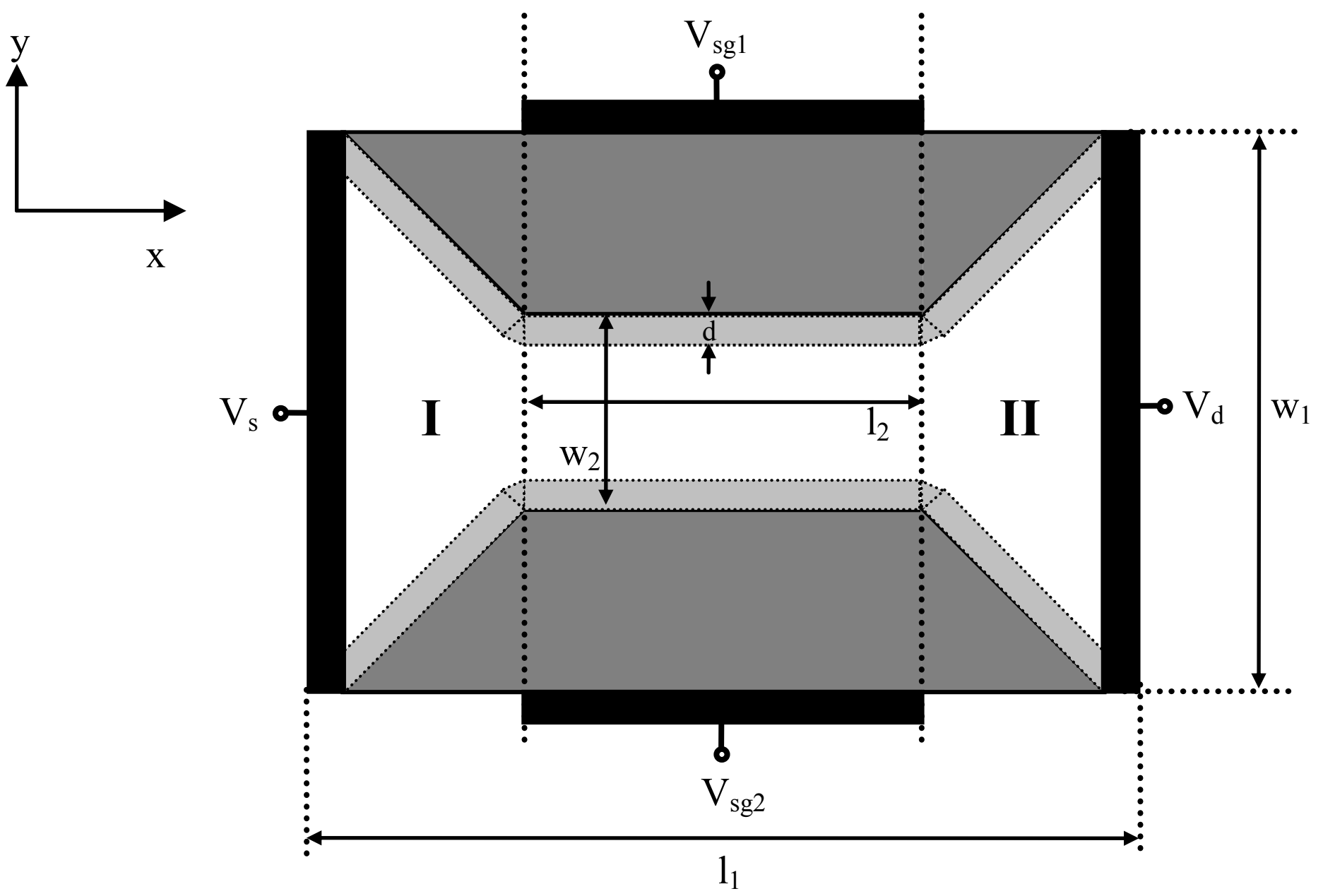

Figure 2

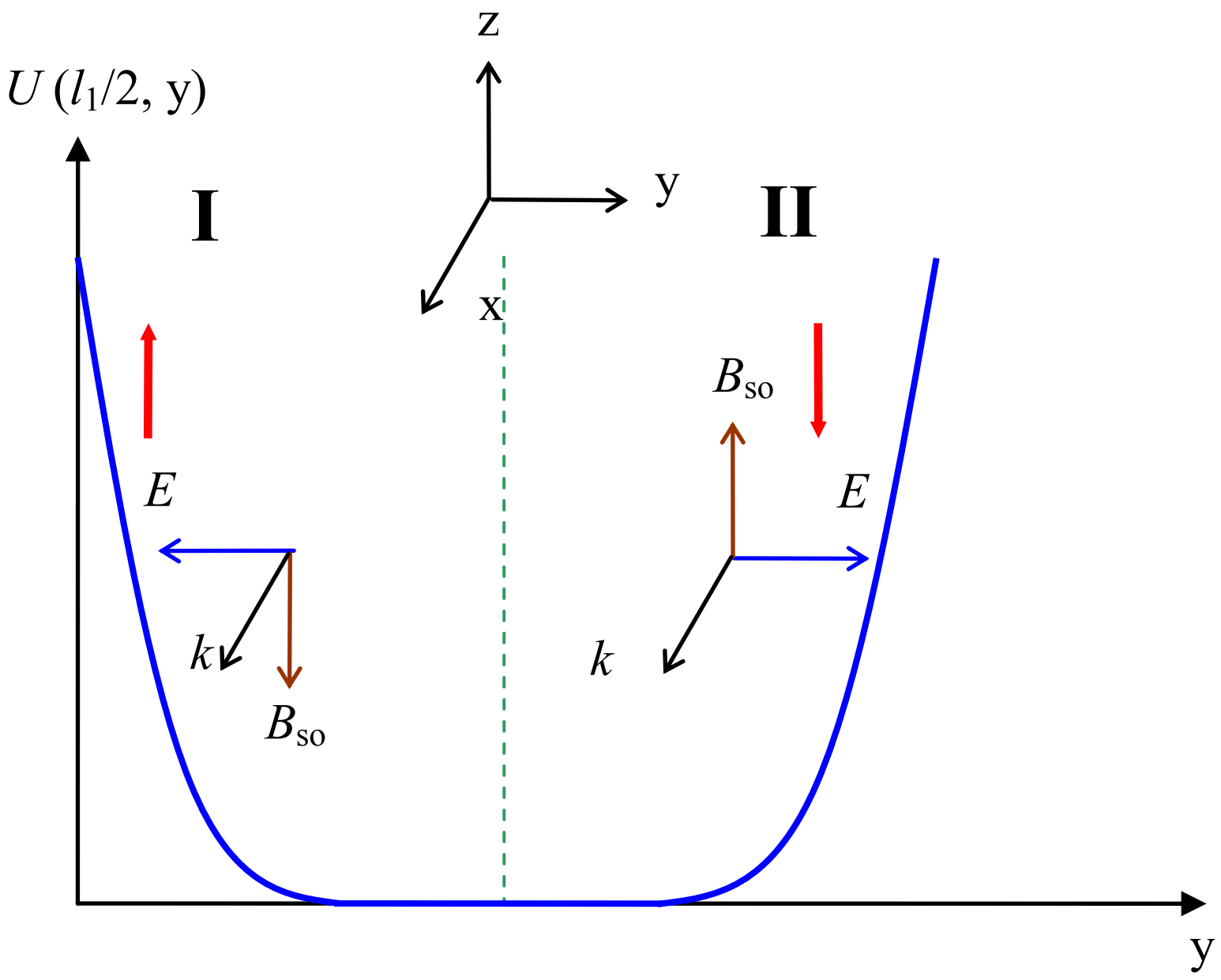

Figure 3(a)

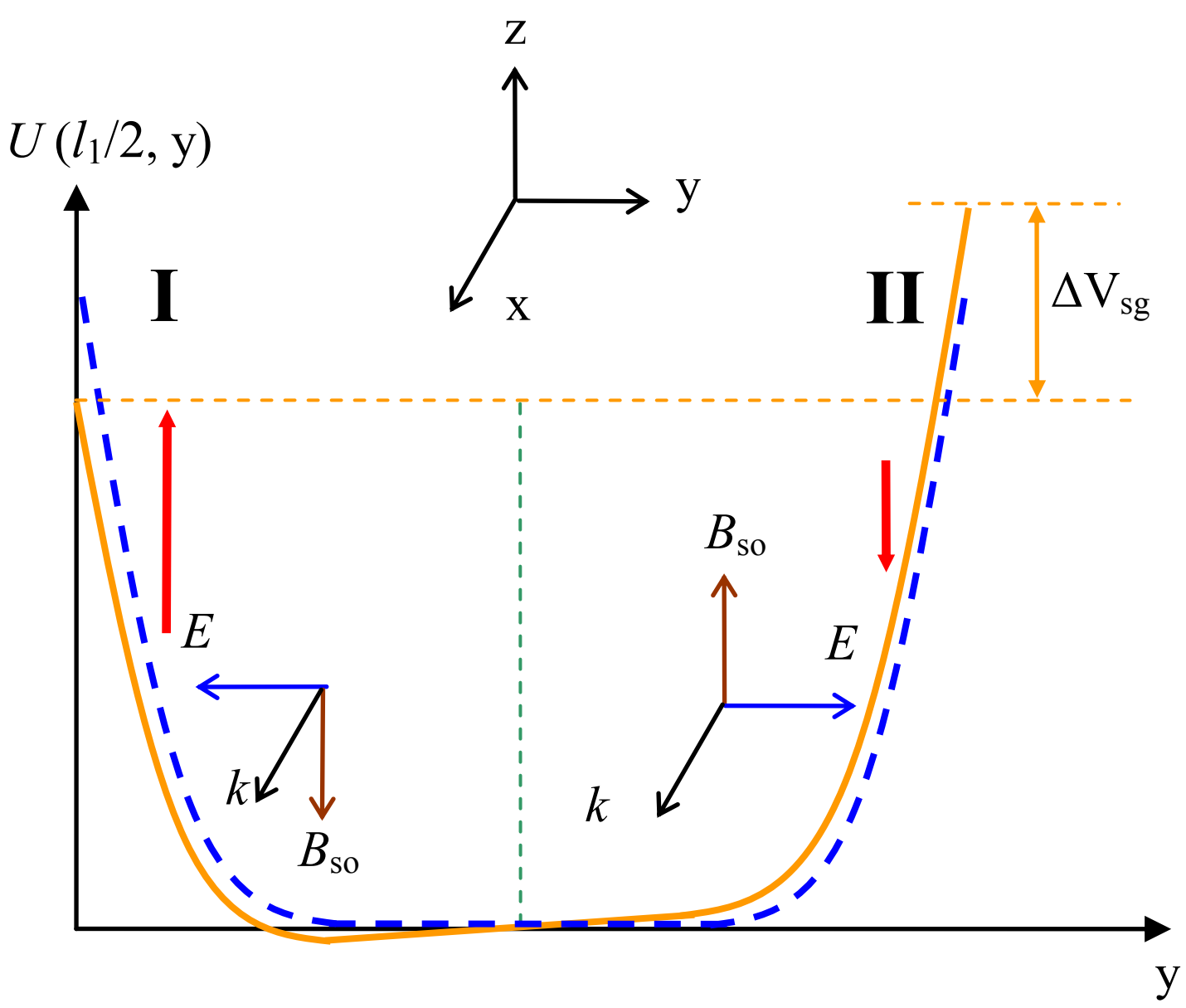

Figure 3(b)

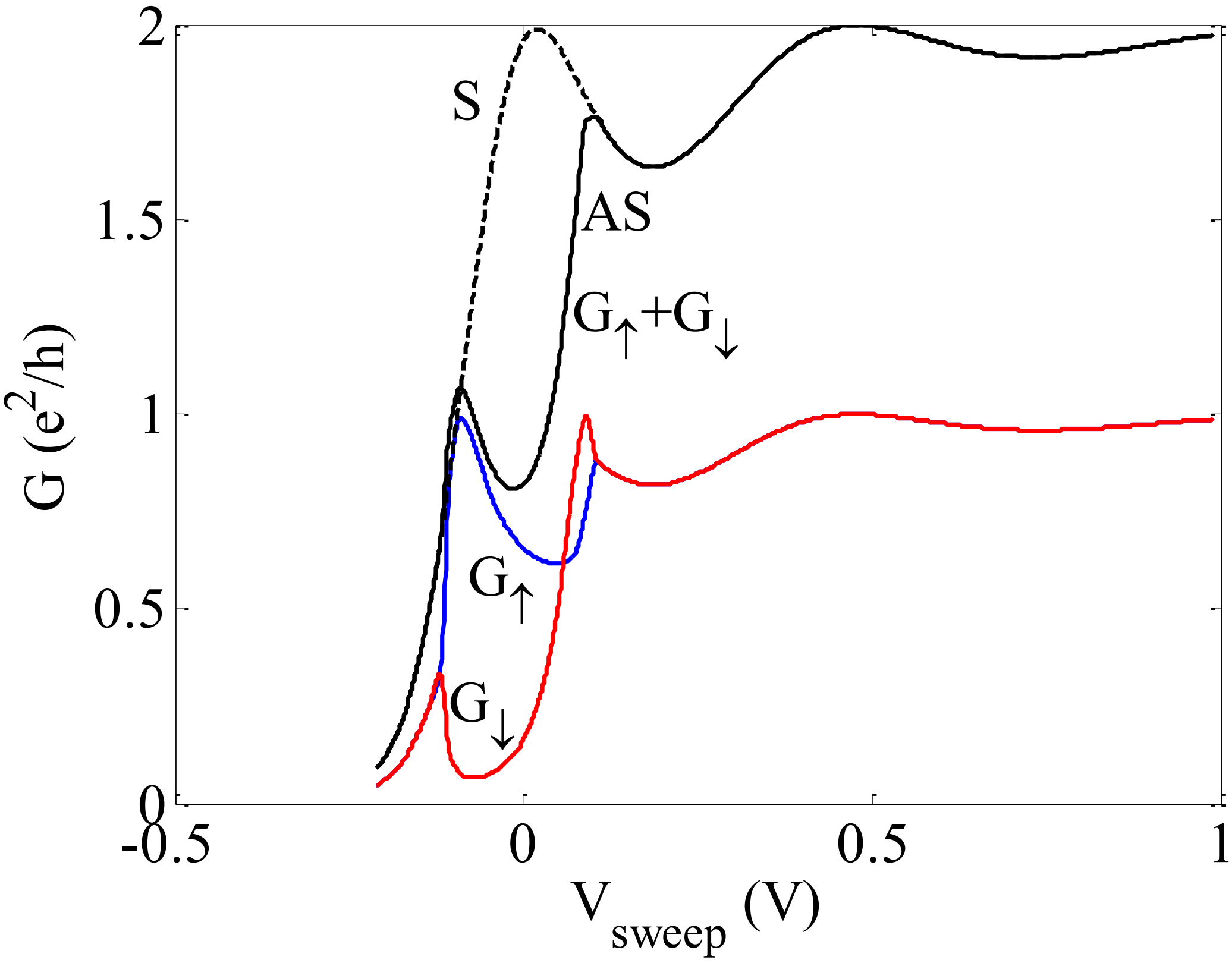

Figure 4

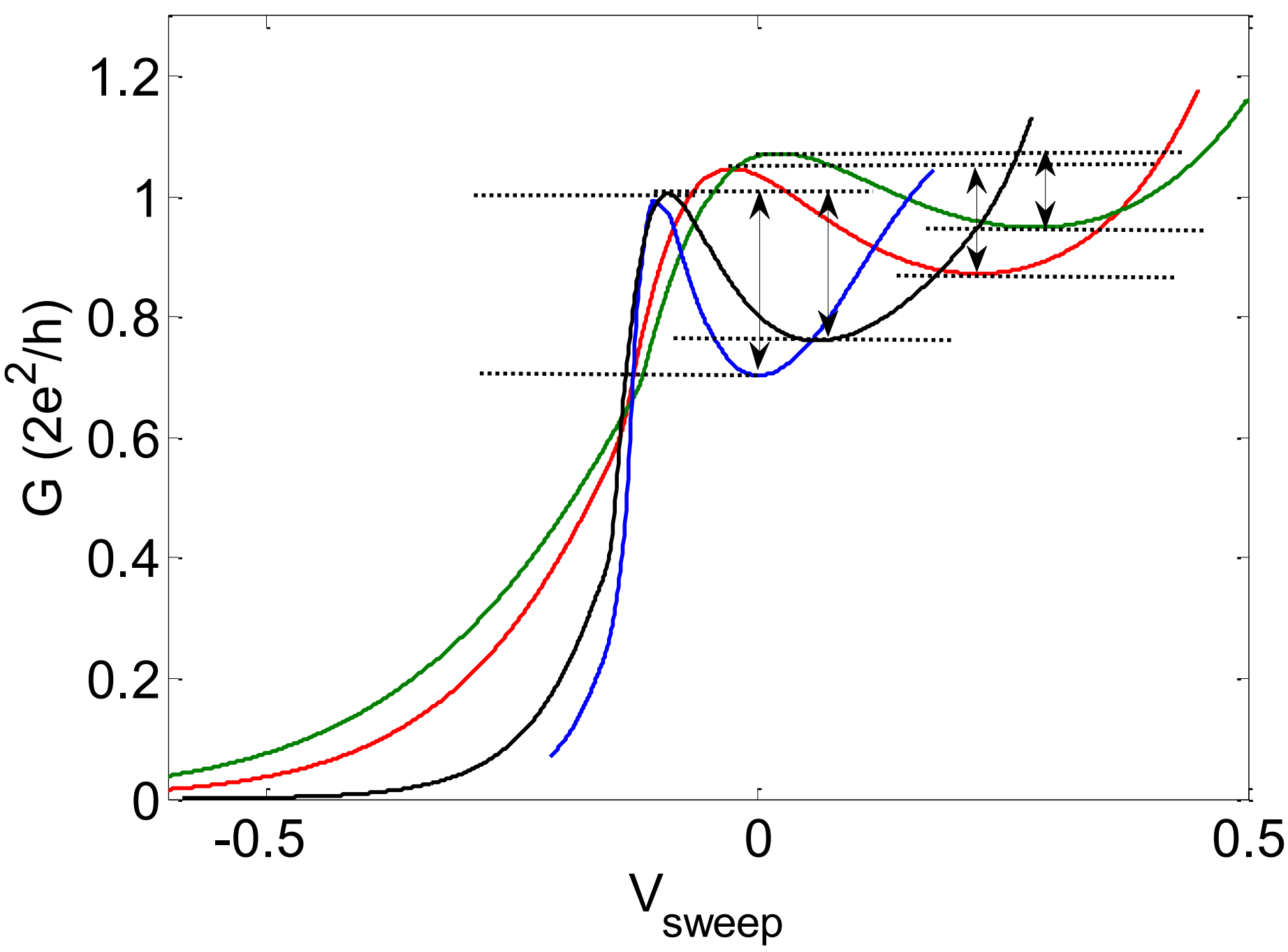

Figure 5

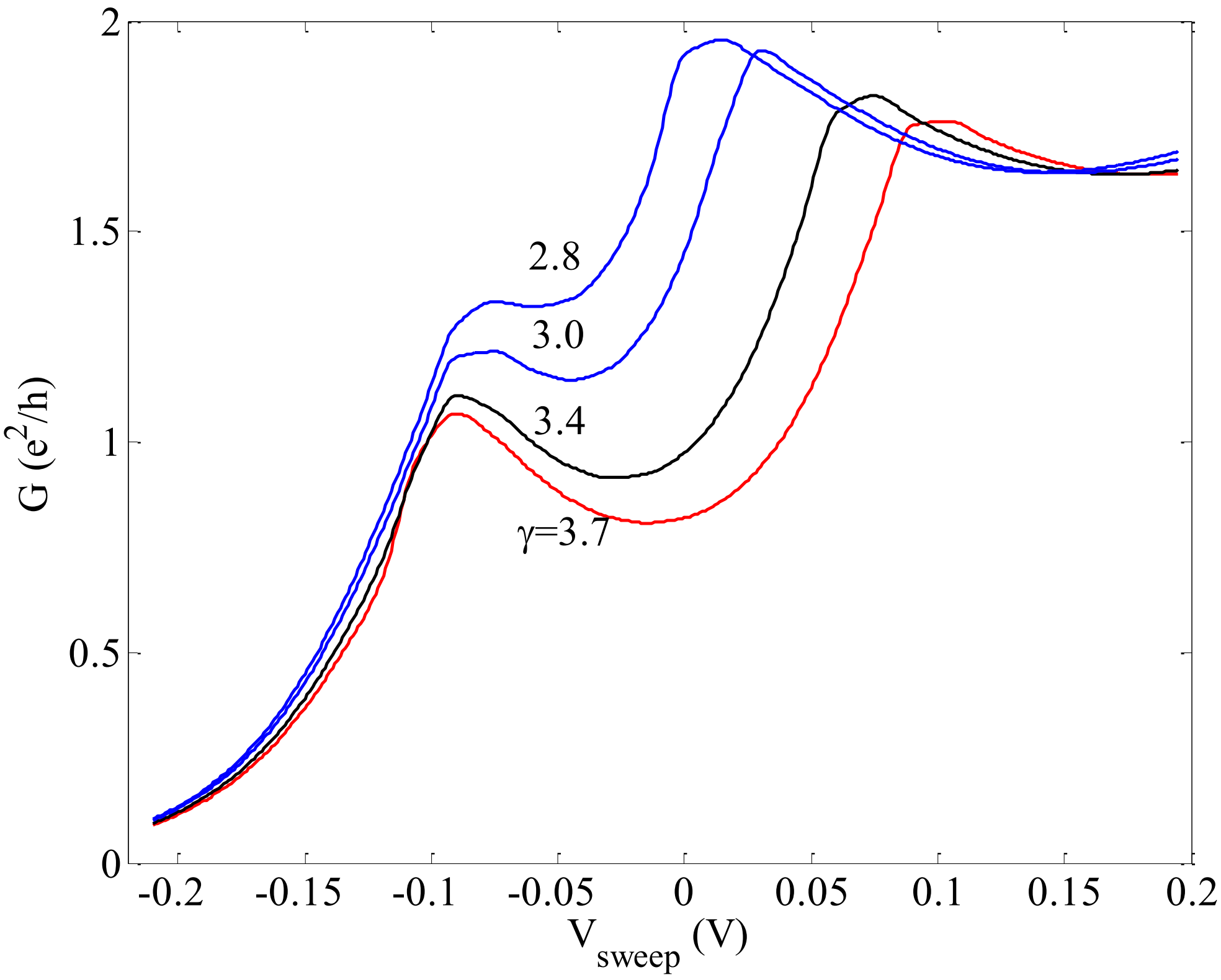

Figure 6

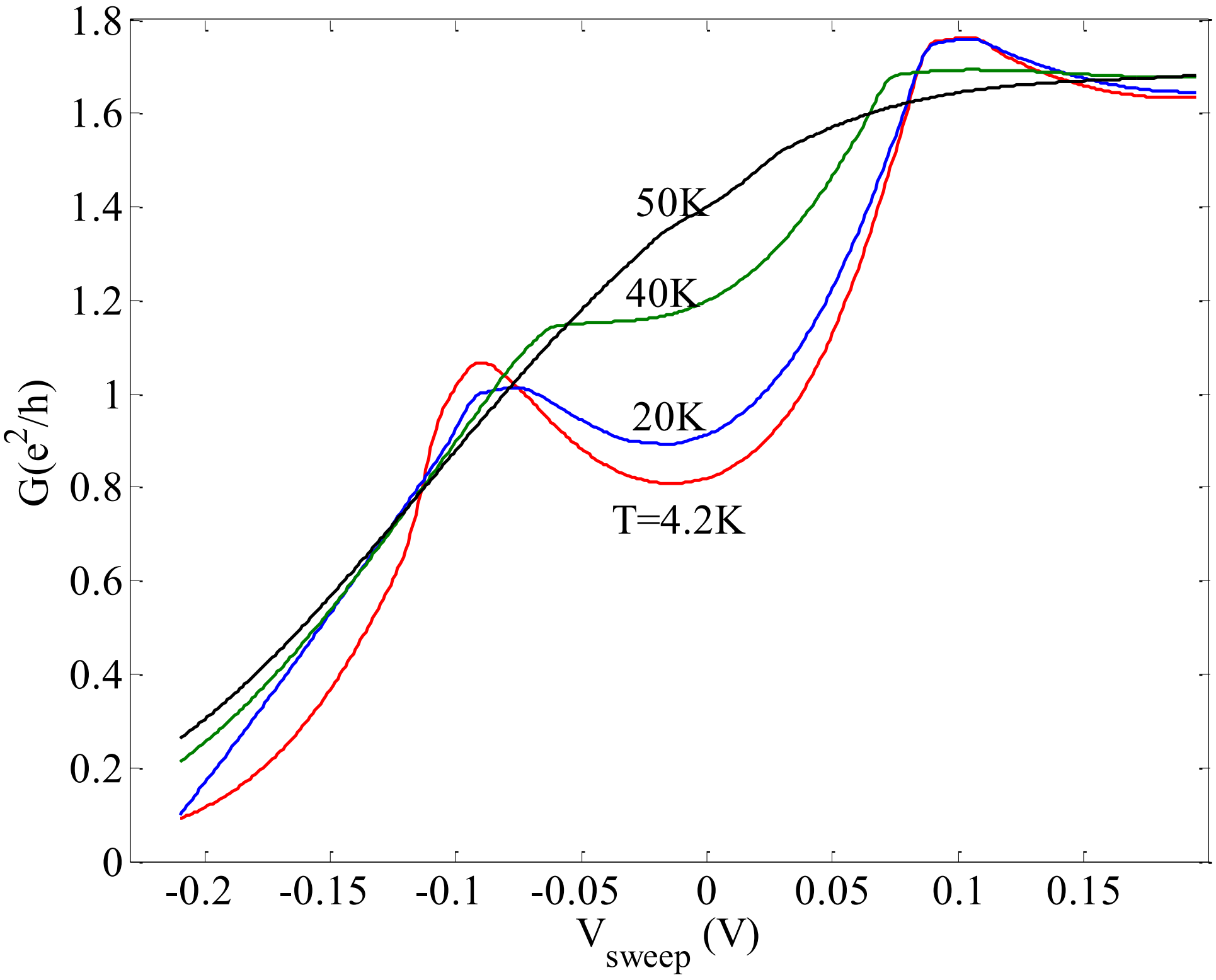

1.8
1.6
1.4
1.2
1
0.8
0.6
0.4
0.2
0
G(e^2/h)
-0.2
-0.15
-0.1
-0.05
0
0.05
0.1
0.15
V_sweep (V)
50K
40K
20K
T=4.2K

Figure 7

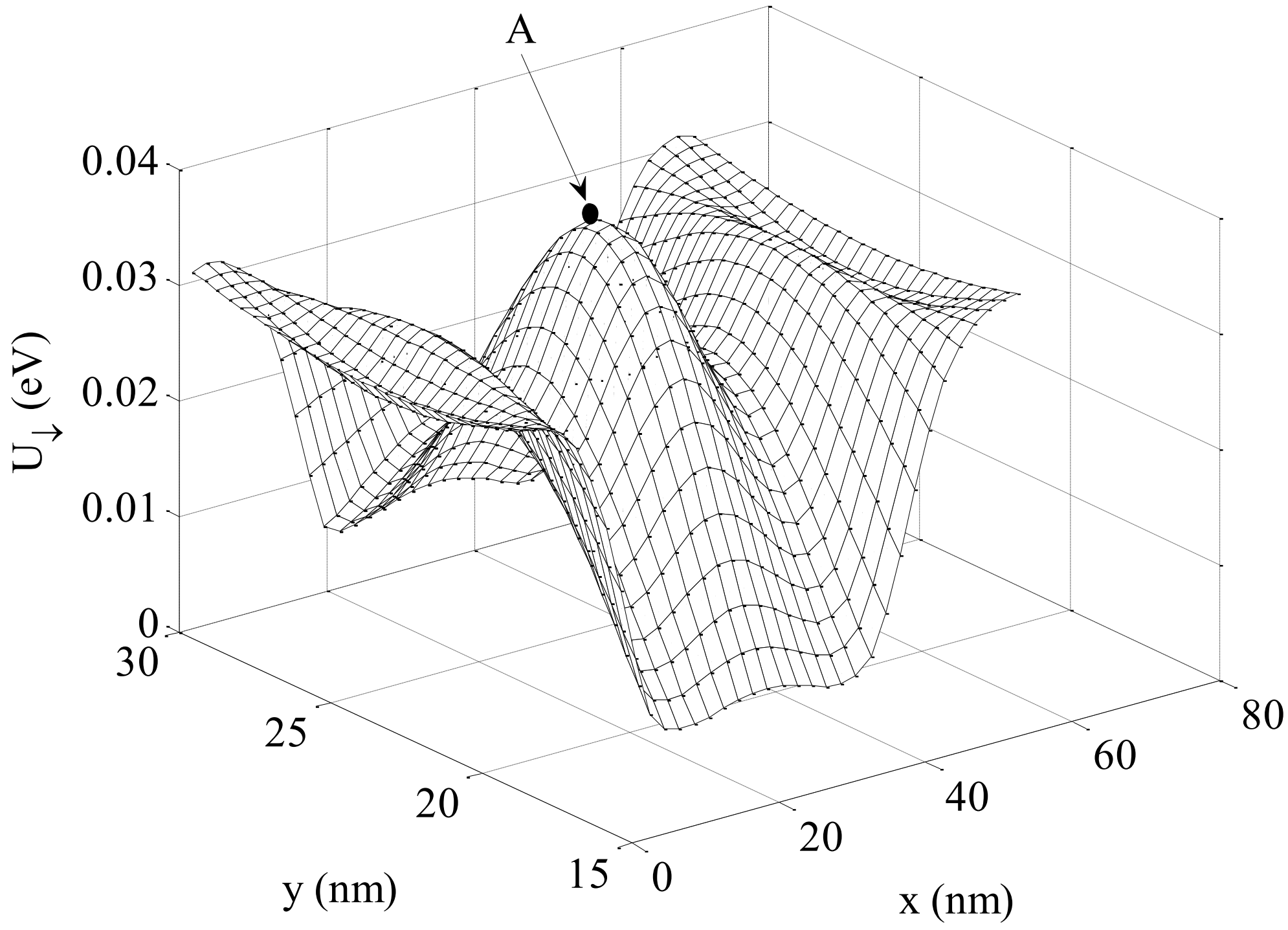

Figure 8(a)

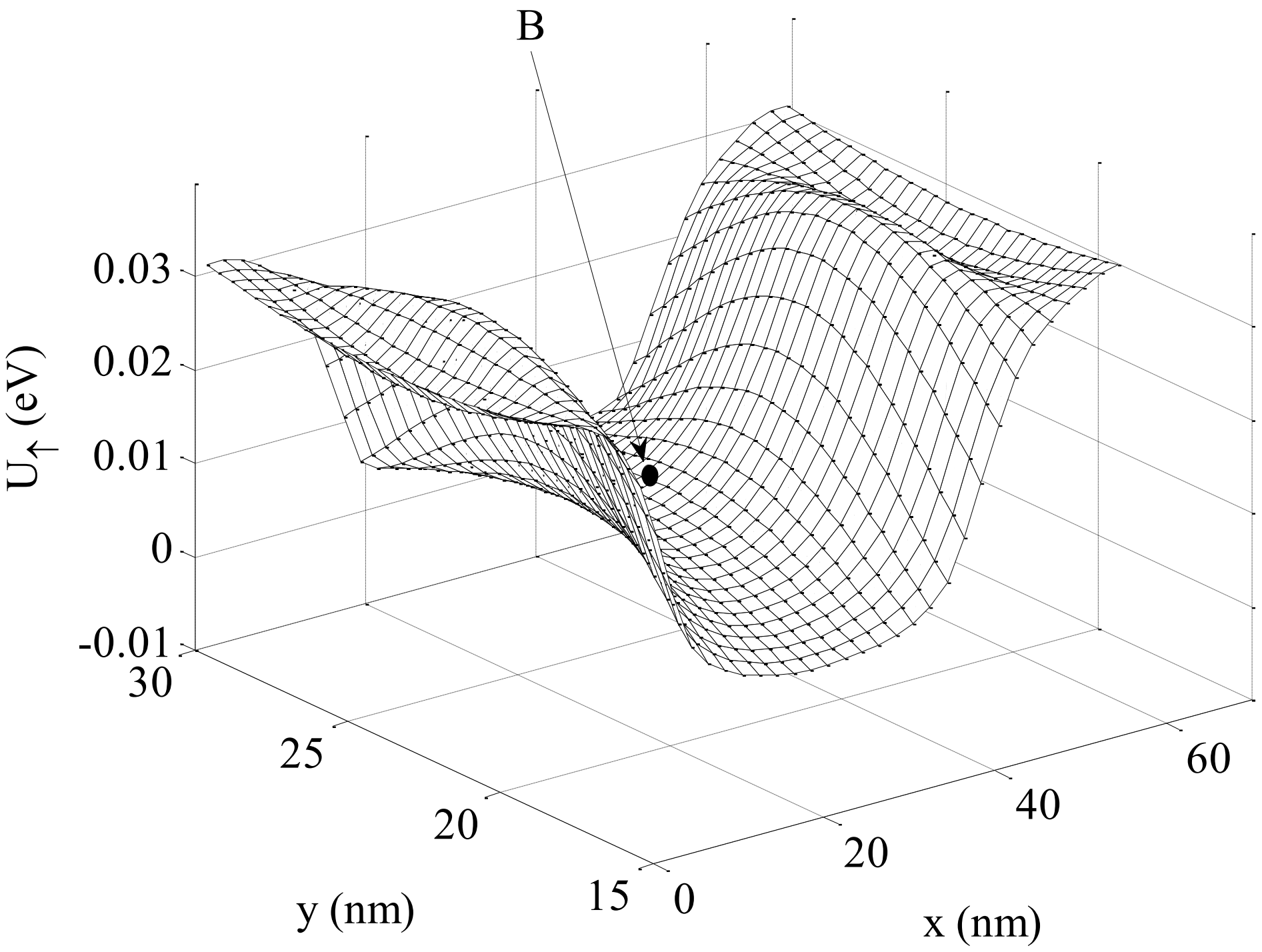

Figure 8(b)

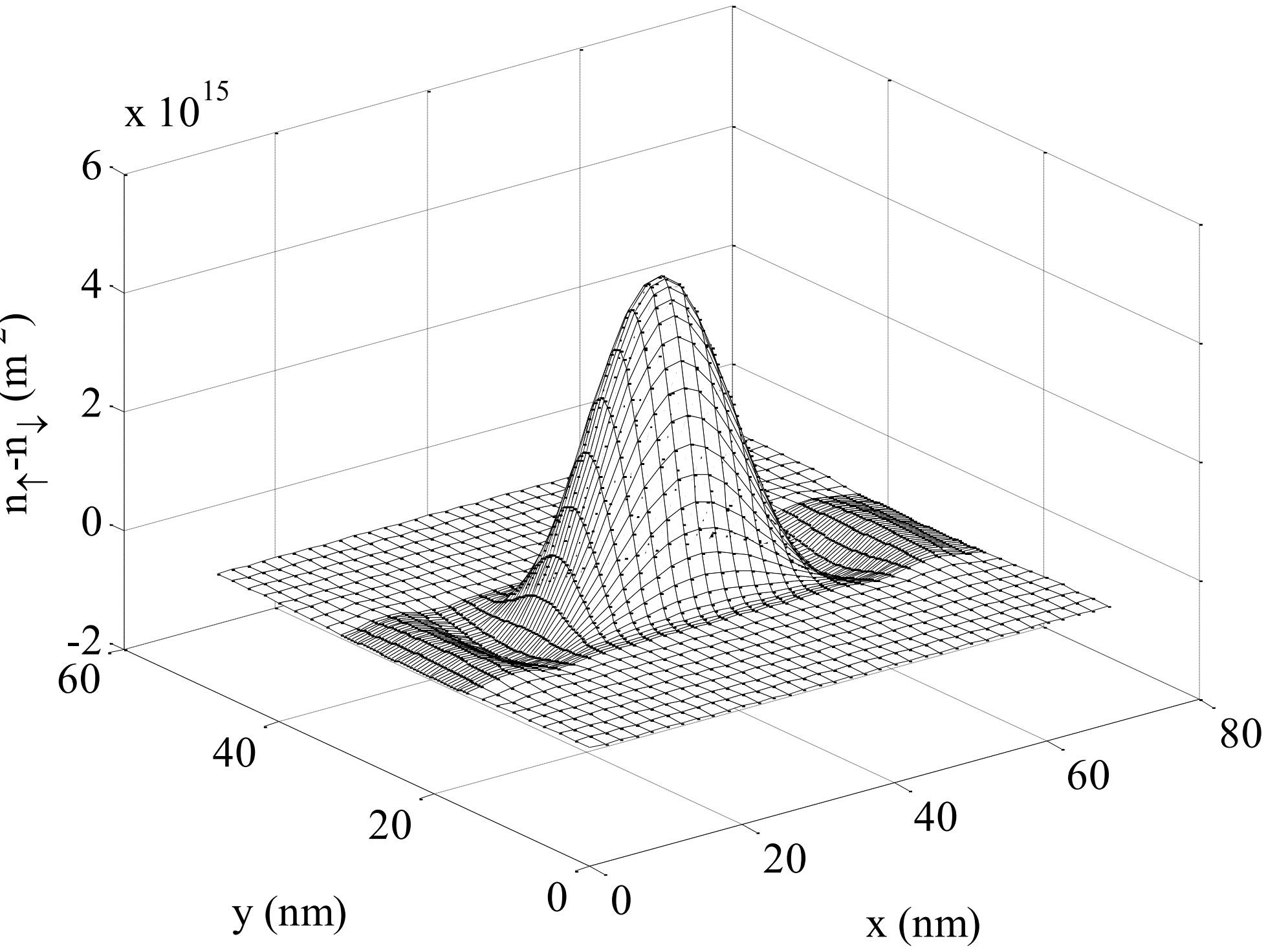

Figure 9

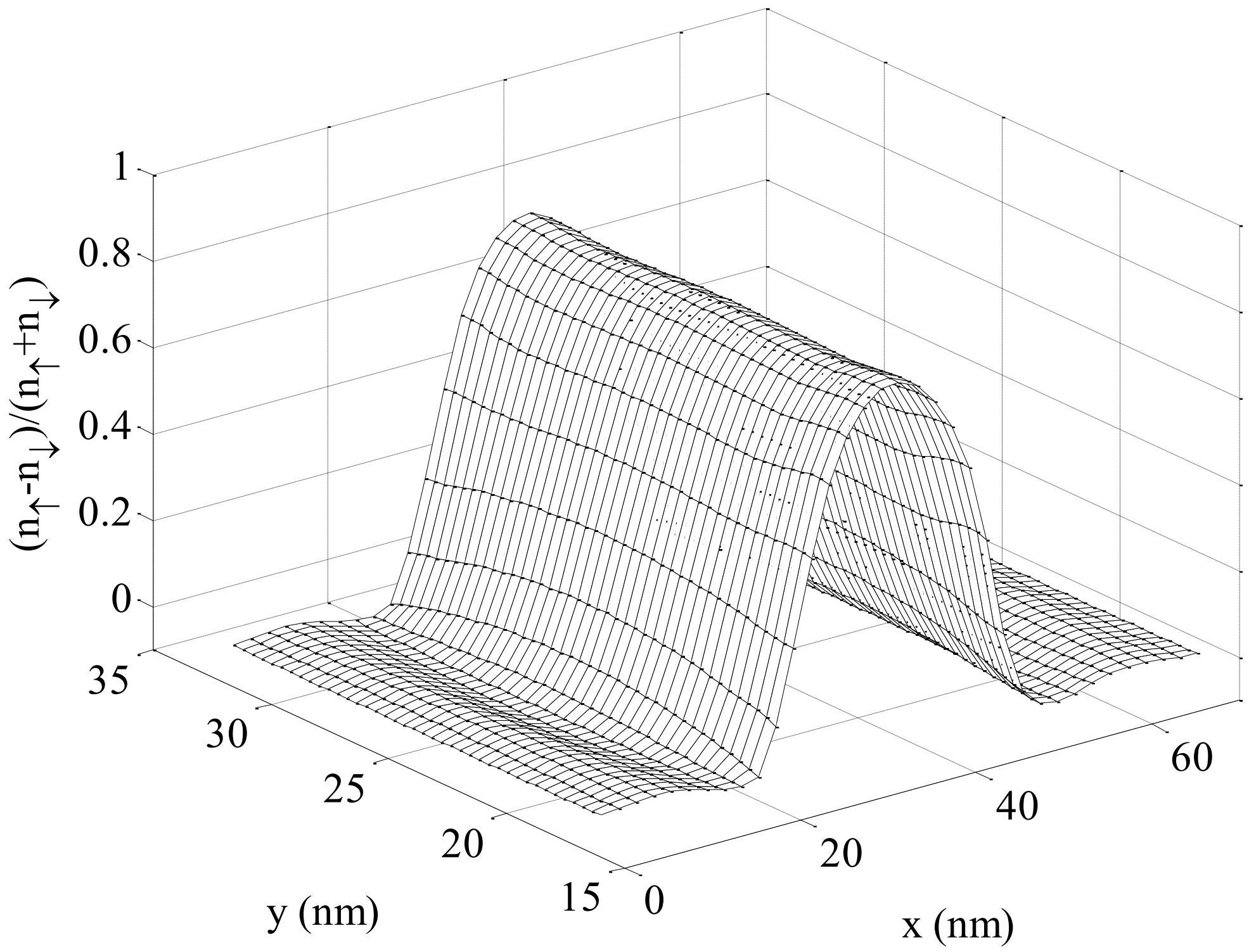

$(n_\uparrow - n_\downarrow)/(n_\uparrow + n_\downarrow)$
1
0.8
0.6
0.4
0.2
0
35
30
25
20
15
y (nm)
0
20
40
60
x (nm)

Figure 10

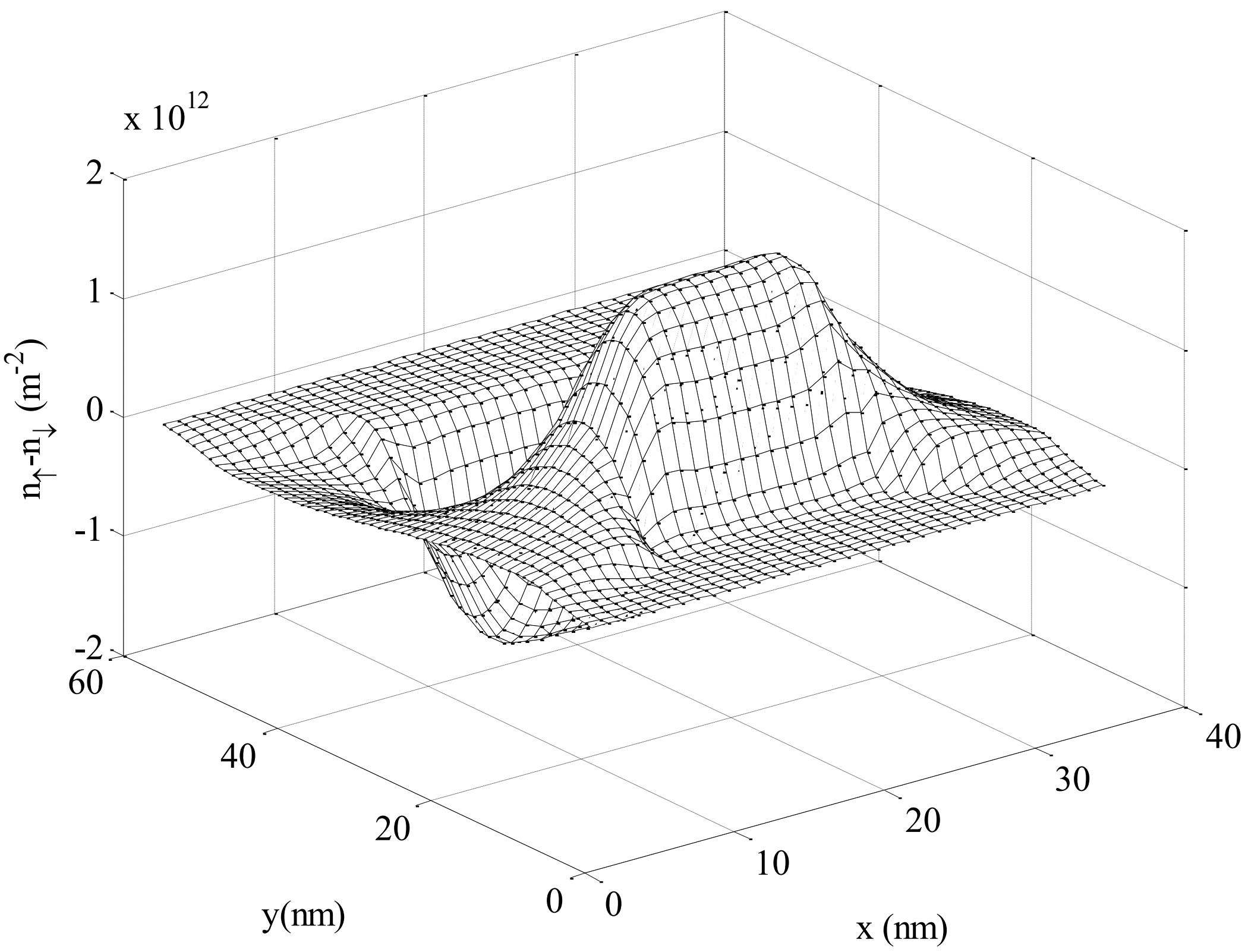
x 10^12
2
1
0
-1
-2
$n_\uparrow$-$n_\downarrow$ ($m^{-2}$)
60
40
20
0
y(nm)
0
10
20
30
40
x (nm)

Figure 11 (a)

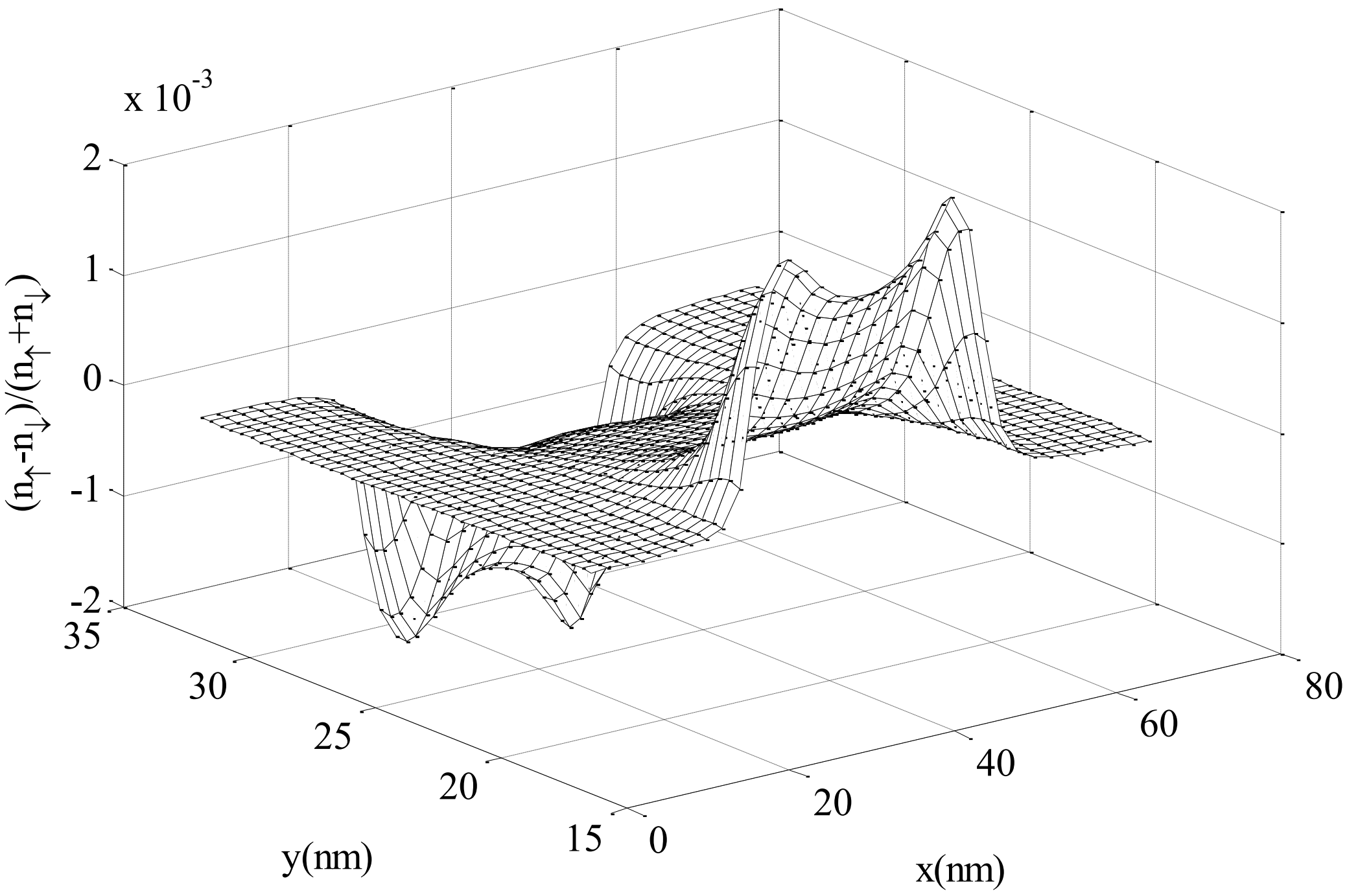

Figure 11 (b)